\documentclass[sigconf]{acmart}
\usepackage{graphicx}
\usepackage{amsmath,amssymb,amsfonts}
\usepackage{algorithmic}
\usepackage{graphicx}
\usepackage{textcomp}
\usepackage{hyperref}
\newcommand{\etal}{\textit{et al}.}
\newcommand{\ie}{\textit{i}.\textit{e}.}
\newcommand{\eg}{\textit{e}.\textit{g}.}
\newcommand{\etc}{\textit{etc}}

\AtBeginDocument{%
  \providecommand\BibTeX{{%
    \normalfont B\kern-0.5em{\scshape i\kern-0.25em b}\kern-0.8em\TeX}}}

\copyrightyear{2020}
\acmYear{2020}
\setcopyright{acmcopyright}\acmConference[MM '20]{Proceedings of the 28th ACM International Conference on Multimedia}{October 12--16, 2020}{Seattle, WA, USA}
\acmBooktitle{Proceedings of the 28th ACM International Conference on Multimedia (MM '20), October 12--16, 2020, Seattle, WA, USA}
\acmPrice{15.00}
\acmDOI{10.1145/3394171.3413928}
\acmISBN{978-1-4503-7988-5/20/10}

\begin{document}
\fancyhead{}

\title{NuI-Go: Recursive Non-Local Encoder-Decoder Network for Retinal Image Non-Uniform Illumination Removal}


%\author{Anonymous Submission, Paper ID:}

\author{Chongyi Li}
\affiliation{%
	\institution{Nanyang Technological University}
}
\email{lichongyi25@gmail.com}

\author{Huazhu Fu}
\authornote{Corresponding author.}
\affiliation{%
	\institution{Inception Institute of Artificial Intelligence (IIAI)}
}
\email{hzfu@ieee.org}

\author{Runmin Cong}
\authornotemark[1]
\affiliation{%
	\institution{Institute of Information Science, Beijing Jiaotong University}
}
\email{rmcong@bjtu.edu.cn}

\author{Zechao Li}
\affiliation{%
	\institution{Nanjing University of Science and Technology}
}
\email{zechao.li@njust.edu.cn}

\author{Qianqian Xu}
\affiliation{%
	\institution{Institute of Computing Technology, Chinese Academy of Sciences}
}
\email{xuqianqian@ict.ac.cn}


\begin{abstract}
Retinal images have been widely used by clinicians for early diagnosis of ocular diseases. However, the quality of retinal images is often clinically unsatisfactory due to eye lesions and imperfect imaging process. One of the most challenging quality degradation issues in retinal images is non-uniform which hinders the pathological information and further impairs the diagnosis of ophthalmologists and computer-aided analysis. To address this issue, we propose a non-uniform illumination removal network for retinal image, called NuI-Go, which consists of three Recursive Non-local Encoder-Decoder Residual Blocks (NEDRBs) for enhancing the degraded retinal images in a progressive manner. Each NEDRB contains a feature encoder module that captures the hierarchical feature representations, a non-local context module that models the context information, and a feature decoder module that recovers the details and spatial dimension. Additionally, the symmetric skip-connections between the encoder module and the decoder module provide long-range information compensation and reuse. Extensive experiments demonstrate that the proposed method can effectively remove the non-uniform illumination on retinal images while well preserving the image details and color. We further demonstrate the advantages of the proposed method for improving the accuracy of retinal vessel segmentation. 
\url{https://li-chongyi.github.io/Proj_ACMMM20_NuI-Go}
\end{abstract}


%\begin{CCSXML}
%<ccs2012>
%  <concept_id>10010520.10010553.10010562</concept_id>
%  <concept_desc>Computer systems organization~Embedded systems</concept_desc>
%  <concept_significance>500</concept_significance>
%  <concept_id>10010520.10010575.10010755</concept_id>
%  <concept_desc>Computer systems organization~Redundancy</concept_desc>
%  <concept_significance>300</concept_significance>
%  <concept_id>10010520.10010553.10010554</concept_id>
%  <concept_desc>Computer systems organization~Robotics</concept_desc>
%  <concept_significance>100</concept_significance>
%  <concept_id>10003033.10003083.10003095</concept_id>
%  <concept_desc>Networks~Network reliability</concept_desc>
%  <concept_significance>100</concept_significance>
%</ccs2012>
%\end{CCSXML}
%
%\ccsdesc[500]{Computer systems organization~Embedded systems}
%\ccsdesc[500]{Computing methodologies~Computer vision problems}

\begin{CCSXML}
<ccs2012>
   <concept>
       <concept_id>10010147.10010178.10010224.10010245</concept_id>
       <concept_desc>Computing methodologies~Computer vision problems</concept_desc>
       <concept_significance>500</concept_significance>
       </concept>
 </ccs2012>
\end{CCSXML}

\ccsdesc[500]{Computing methodologies~Computer vision problems}

\keywords{retinal image enhancement; non-uniform illumination removal; deep learning}



\maketitle

\section{Introduction}
\label{sec:introduction}
\vspace{0.3cm}
Retinal images play important roles in diagnosing and monitoring retinal diseases in early stages, including glaucoma, diabetic retinopathy, pathological myopia, \etc. \cite{Abramoff2010,Chuaudhuri1989,Li2004survey}. However, retinal images often suffer from the issues of non-uniform illumination, low illumination, low contrast, and detail blurring due to imperfect imaging process \cite{Swanson1993, Hani2009,Zhou2018}. Such quality degraded retinal images significantly impair the diagnosis of ophthalmologists and also affect the performance of computer-aided analysis of retinal diseases \cite{Cheng2018}, such as retinal vessel segmentation, optical disc segmentation, and vascular structure recognition. One of the most challenging quality degradation issues in retinal images is the non-uniform illumination which is the main focus in this paper. It is necessary to remove the non-uniform illumination on retinal images before using them in diagnosis.

Due to the limited generalization capability, the traditional image enhancement methods (\eg, Histogram Equalization (HE), Contrast Limited Adaptive Histogram Equalization (CLAHE), and Gamma correction) that either adjust the histogram distribution of images or globally correct the pixel values cannot be directly extended to retinal images. Sometimes, these traditional methods even tend to introduce unfriendly artifacts when they are used to process the low-quality retinal images. These artifacts severely affect the accuracy of diagnosis, which cannot be used in practical conditions. To this end, several image enhancement methods specially designed for retinal images have been proposed ~\cite{Zhou2018,Cheng2018,Feng2007,Saha2018}. Unfortunately, the performance of retinal image enhancement remains largely unsatisfactory, especially in practical applications, such as vessel segmentation, vascular structure recognition, and disease diagnosis. In addition, recent years have witnessed the significant success of deep learning in medical image processing \cite{Shao2019,Yan2019,Jiang2019,Fu2018,Moreau2018}. However, deep learning-based retinal image enhancement receives less attention due to lacking sufficient training data and effective network architectures.
In this paper, we will address these issues and provide the first deep learning-based solution for non-uniform illumination removal in retinal images.

\begin{figure}[!t]
	\centering
	\centerline{\includegraphics[width=0.4\textwidth]{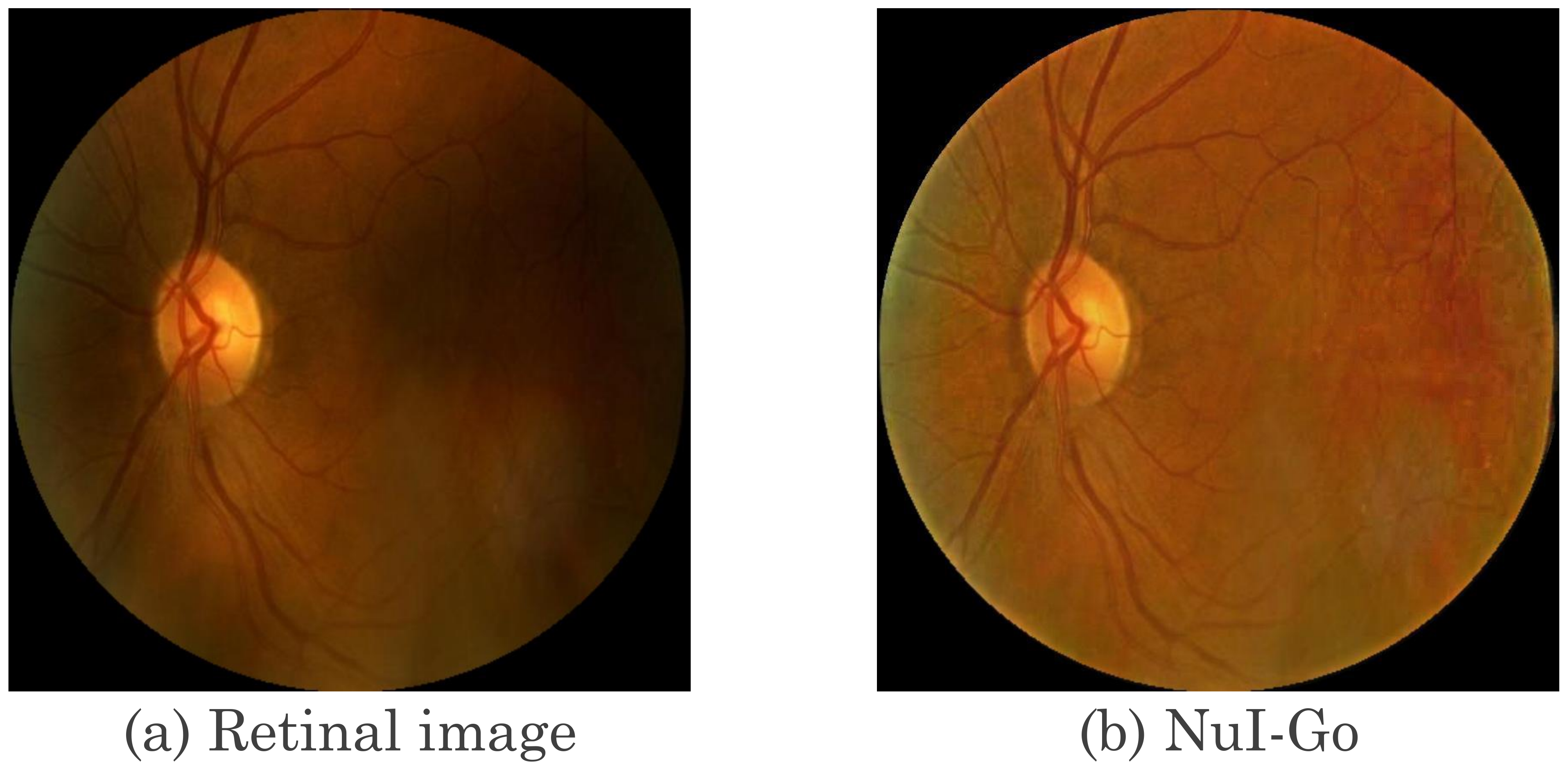}}
	\caption{Sample result on a real retinal image. The NuI-Go can correct the illumination of retinal image while well preserving its details and color.}
	\label{fig:im_sample}
\end{figure}

Different from existing methods, our method combines the capacity of neural networks with image physical formation model to study the issue of retinal image non-uniform illumination, which leads to more reasonable and reliable enhancement performance.  Concretely, we first approximate the retinal image formation model with non-uniform illumination by combining the domain knowledge of retinal images with the human-lens attenuation and scattering model. With the synthetic retinal images with degradations, a non-local encoder-decoder recursive network architecture is proposed to learn the mapping between the retinal image with non-uniform illumination and the corresponding clear counterpart. The degraded retinal image is enhanced by recursive Non-local Encoder-Decoder Residual Blocks (NEDRBs) in a progressive manner. Extensive experiments demonstrate that the proposed method achieves the state-of-the-art performance on both real and synthetic images in qualitative and quantitative metrics. In addition, we further demonstrate the advantages of the proposed method for subsequent retinal vessel segmentation task. In Figure~\ref{fig:im_sample}, we present a sample result of our NuI-Go. Our NuI-Go can effectively correct the illumination of the retinal image while well preserving its detail and color information.

The contributions of this work are as follows.
\begin{itemize}
  \item This work is the first attempt to provide a deep learning solution for the problem of retinal image non-uniform illumination removal, which can effectively correct the illumination of a retinal image while well preserving the original details and color appearances.
  \item We propose a simple yet effective non-local encoder-decoder network to progressively enhance degraded retinal images, which integrates local and non-local information simultaneously.
  \item Benefiting from novel network architecture and physical model-based non-uniform illumination synthesis strategy, our method achieves impressive performance on retinal image non-uniform illumination removal.
\end{itemize}

\section{Related Work}
\vspace{0.3cm}
In this paper, we mainly focus on image light enhancement methods that can be categorized into traditional image enhancement methods and retinal image enhancement methods.

\noindent
\textbf{Traditional Image Enhancement Methods.}
HE-based methods~\cite{Coltuc2006,Ibrahim2007,Celik2011,Stark2000,Lee2013,Li2016,LiPRL17} adjust the histogram distribution of image to improve its contrast. However, HE-based methods tend to over-enhance the images due to impertinently changing the dynamic range of image and some local HE methods are time-consuming.
Unlike the HE-based methods, the Retinex theory~\cite{Land1977} assumes that the image can be decomposed into reflectance and illumination, and has been widely used in low-light image enhancement thanks to its physical interpretability.
Generally, the reflectance component is treated as the enhanced image \cite{Li2014,Rother2011,Wang2013,Fu2016,Guo2017,Li2018}.
However, there are some limitations of Retinex-based image enhancement due to the gaps between real-world high-quality images and the reflectance component. Thus, the results of Retinex-based methods look unnatural and unrealistic.
In addition, the last decade has witnessed an increasing interest in image light enhancement \cite{Lore2017,Chen2018,LiPRL2018,Ren2019,Wang2019,LiPR2020,LiZero2020}, especially with the emergence of deep learning.

\noindent
\textbf{Retinal Image Enhancement Methods.} Recently, some methods specially designed for retinal image enhancement have been proposed.
Cheng \etal~\cite{Cheng2018} proposed a structure-preserving guided retinal image filter, called SGRIF.
Feng~\etal~\cite{Feng2007} used the Coutourlet transform to achieve retinal image contrast enhancement.
In \cite{Mitra2018}, a non-uniform illuminated fundus image enhancement method was proposed, which reduces the blurriness of fundus images based on the cataract physical model and enhances the images with an objective on contrast perfection with no preamble of artifacts.
Saha \etal~\cite{Saha2018} proposed a two-step method to process the non-uniform or poor illumination of retinal images, including an illumination correction algorithm and a color restoration algorithm.
Zhou \etal~\cite{Zhou2018} designed a retinal image enhancement method based on luminosity and contrast adjustment.

Despite the prolific work, the performance of retinal image enhancement is still far from satisfactory and practical. With the rise of deep learning technology~\cite{Deng2009,Residual,LiSPL18,GuoTIP18,LiTGARS19}, we see a new dawn in performance improvement; however, deep learning-based retinal image enhancement has not been fully explored. To the best of our knowledge, the proposed NuI-Go network is the first deep learning-based non-uniform illumination removal method for retinal images. Besides, the proposed retinal image non-uniform illumination synthesis strategy can be used for deep models training and also for the evaluations of retinal image enhancement methods.

\section{Proposed Method}
\vspace{0.3cm}
%\section{Proposed Method}
\begin{figure*}[!t]
\centering
\centerline{\includegraphics[width=1.\textwidth]{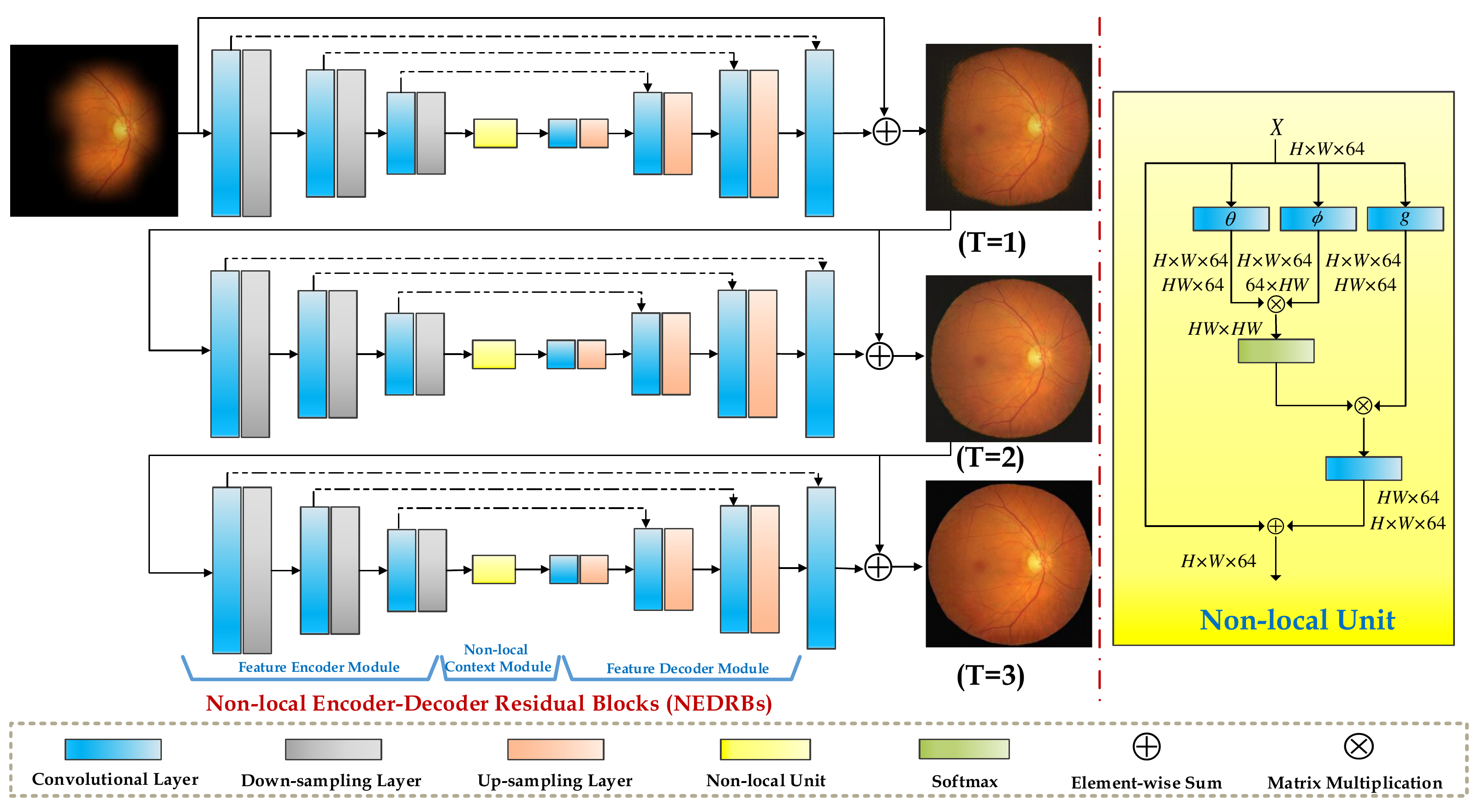}}
\caption{The network architecture of NuI-Go. \textbf{Left side}: three recursive NEDRBs, where each NEDRB contains a feature encoder module, a non-local context module, and a feature decoder module.  \textbf{Right side}: the details of non-local unit, where ``$H\times W\times64$'' denotes height$\times$width$\times$64 channels of feature maps $X$ (reshaping is performed when noted).}
\label{fig_1}
\end{figure*}

The network architecture is presented in Figure~\ref{fig_1}. As illustrated, the NuI-Go network consists of three recursive non-local encoder-decoder residual blocks (NEDRBs) that share the same structure. It takes a degraded retinal image as input and produces a set of enhanced results in a coarse-to-fine scheme. The last output is treated as our final enhanced result. In what follows, we explain the NuI-Go network structure, the loss functions, and the non-uniform illumination retinal image synthesis approach in detail.

\subsection{Proposed NuI-Go Network}
\begin{figure}[!t]
\centering
\centerline{\includegraphics[width=0.48\textwidth]{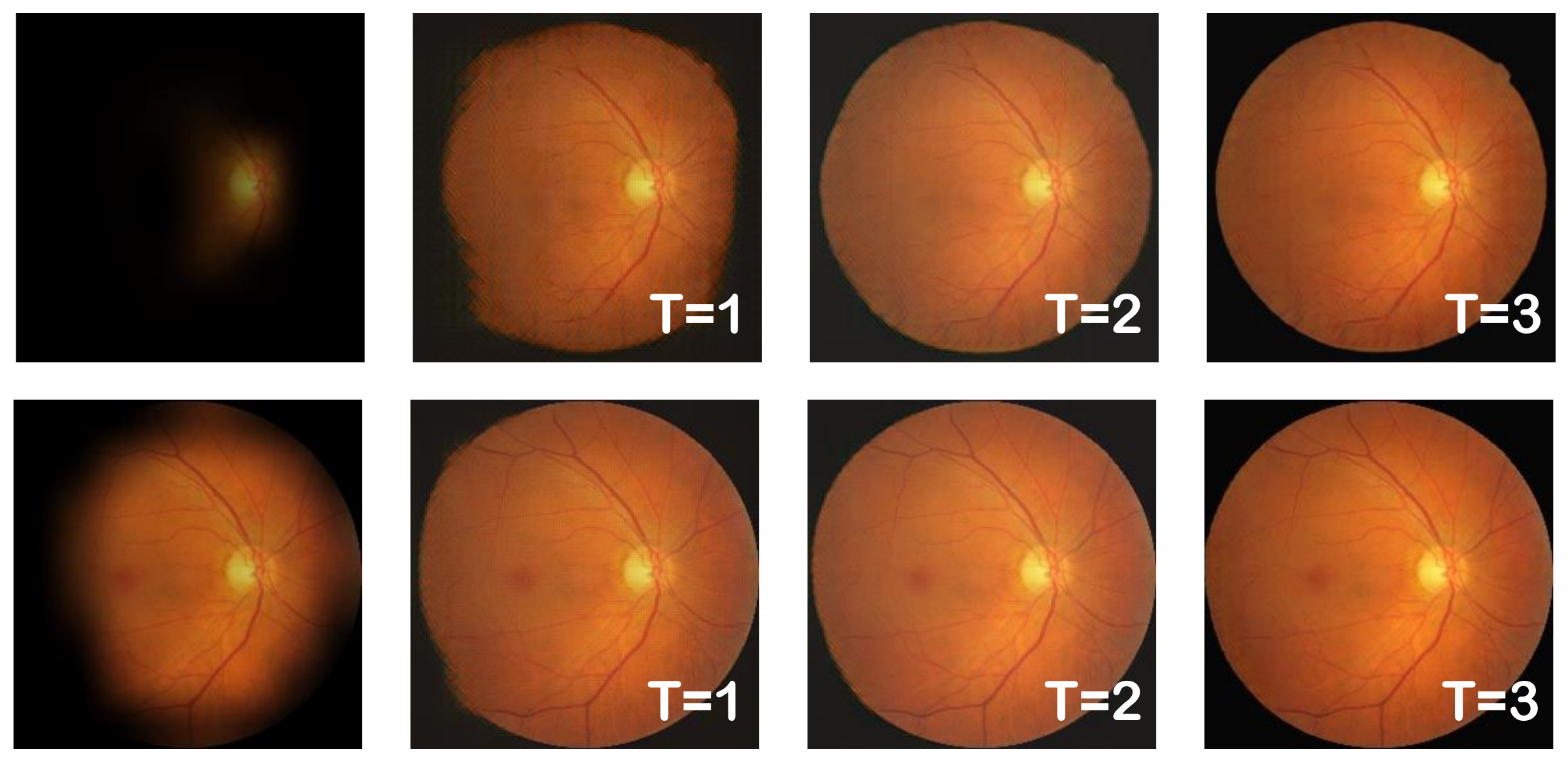}}
\caption{Stage-wise results of the NuI-Go network. From left to right are the input retinal images, the results of stage T=1, 2, and 3, respectively.}
\label{fig:stage-wise}
\end{figure}

\vspace{0.3cm}
We suggest tackling the retinal image non-uniform illumination problem in multiple stages, where a simple yet effective recursive non-local encoder-decoder residual blocks (NEDRB) is used at each stage. Such a recursive manner has several advantages than deeper and more complex networks:
\begin{enumerate}
  \item it is computationally efficient;
  \item it is easy to be trained; and
  \item it allows to stop the inference at any stage in practical applications (i.e., our method provides a trade-off between inference time and accuracy).
\end{enumerate}
Note that, the performance gains by adding more stages (T$>$3) will become marginal, but at the cost of more training and inference time. Thus, we choose the third stage's result as our final result. The stage-wise results are provided in Figure~\ref{fig:stage-wise}. As shown, the retinal images with non-uniform illumination are gradually becoming clear after stage-by-stage enhancement. Next, we will explain the proposed NEDRB.

NEDRB enjoys an encoder-decoder structure with an embedded non-local unit, which contains a feature encoder module that captures the hierarchical feature representations, a non-local context module that models the context information, and a feature decoder module that recovers the details and spatial dimension. More specifically, the input image is first fed to the feature encoder module that is built by three convolutional layers followed by 2 $\times$ down-sampling operation:
\begin{equation}
\label{equ_1}
F_{enc}=g(f_{3}(g(f_{2}(g(f_{1}(I))_{\downarrow}))_{\downarrow}))_{\downarrow},
\end{equation}
where $f$ and $g$ are the notation for the convolution and ReLU actication function, respectivaly, $I$ is the input image, $\downarrow$ denotes the 2 $\times$ down-sampling operation. After that, the encoder features $F_{enc}$ is forwarded to the non-local context module, which can be expressed as:
\begin{equation}
\label{equ_2}
F_{nlu}=NLU(F_{enc}),
\end{equation}
where $NLU$ represents the operations in the non-local unit. The retinal image shares structured texture and similar color, which provides a strong prior for correcting the non-uniform illumination regions by the context. Therefore, instead of employing the multi-scale convolution to extract local context information used in traditional encoder-decoder networks, we incorporate the non-local unit \cite{Wang2018} into the proposed encoder-decoder structure, which enlarges the receptive field to the entire image. Our intuition is to incorporate non-local features into local encoder-decoder features for building richer feature representations that combine both non-local and local information. As shown in Figure~\ref{fig_1}(right), the non-local unit computes the response at a position as weighted sum of the features at all spatial positions. 

Specifically, it can be implemented by simple convolution, softmax, \etc. (for embedded Gaussian version's non-local unit), where the unknown parameters can be adaptively learned. Embracing the benefit of non-local context information which captures the long-range feature correlation, our network can effectively focus on the non-uniform illumination regions and predict the potential color and details by the long-range dependencies (context modeling).

At last, the features $F_{nlu}$ are passed through decoder module. Coupled with the symmetric skip connections, the operations in decoder module can be expressed as:
\begin{equation}
\begin{split}
\label{equ_3}
R=I \dotplus &f_{7}(\{f_{6}(\{g(f_{5}(\{g(f_{4}(F_{nlu}))_{\uparrow}\\
&,g(f_{3})\}))_{\uparrow},g(f_{2}\})))_{\uparrow},g(f_{1})\}),
\end{split}
\end{equation}
where $R$ represents the reconstructed result,  $\dotplus$ denotes the element-wise addition, $\uparrow$ is the 2$\times$ up-sampling operation, and $\{,\}$ stands for the concatenation operation along the channel dimention. In our recursive structure, we repeat  the non-local encoder-decoder residual block three times.

In summary, the encoder-decoder structure can fully exploit hierarchical feature representations and reconstruct the enhanced results while the non-local unit captures long-range dependencies for spatial context modeling. Additionally, the symmetric skip-connections provide long-range information compensation and reuse. Finally, we enforce the NEDRB to learn the difference between the degraded retinal image and its ground truth (residual learning \cite{Residual}), which facilitates gradient back-propagation and pixel-wise correction.

\begin{figure*}[!htb]
\centering
\includegraphics[width=18cm,height=3cm]{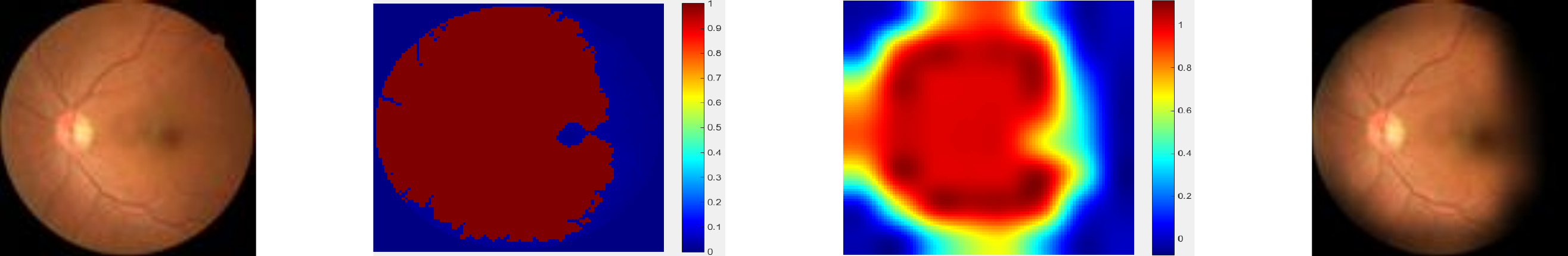}
\caption{An example of the non-uniform illumination synthesis. From left to right are the well-lit retinal image, the coarse illumination mask, the smoothed illumination mask, and the synthetic retinal image. Different color represents different values in the illumination masks that are represented by heatmaps.}
\label{fig_2}
\end{figure*}

In the NuI-Go, the down-sampling layer is implemented by 2$\times$ max pooling while the up-sampling layer is implemented by transposed convolution with the kernel size of 3$\times$3 and stride 2. All convolutional layers have the kernel size of 3$\times$3 and output 64 feature maps, except for the convolutional layers before the results, where each convolutional layer has the kernel size of 1$\times$1 and outputs 3 residual maps. We pad zeros before each convolutional layer to keep the results having the same size as the input image. Moreover, the convolutional layers are followed by ReLU non-linear unit, except for the convolutional layers before the results. For the embedded Gaussian version's non-local unit, we follow the implementation of \cite{Wang2018}. For the computational efficiency, each convolutional layer in the non-local unit has the kernel size of 1$\times$1.

\subsection{Loss Function}
\vspace{0.3cm}
The perceptual loss has demonstrated much better visual performance than the per-pixel losses. However, it usually fails to achieve good quantitative performance~\cite{Johnson2016}. To remit this issue, we incorporate the $\ell_{1}$ loss into our network optimization, which ensures the results have good quantitative scores. We only add the $\ell_{1}$ loss to stage T=3 while the stages T=1 and T=2 are optimized by only using the perceptual loss because it is hard to achieve the convergence when putting these two losses on each stage's output. Besides, such an optimization scheme saves the training time.

We employ the perceptual loss based on the VGG-19 network $\phi$ \cite{Simonyan2015} pre-trained on the ImageNet dataset \cite{Deng2009}. Let $\phi_{j}(x)$ be the $j$th convolutional layer. We measure the distance between the feature representations of the enhanced retinal image $\hat{R}$ and the ground truth image $R$ as:
\begin{equation}
\label{equ_1}
L_{per}^{3}=\sum\limits_{i = 1}^M \mid \phi_{j}(\hat{R_{i}})-\phi_{j}(R_{i})\mid,
\end{equation}
where $M$ is the number of each batch in the training procedure. Similar to the $L_{per}^{3}$ for T=3, the losses of T=1 and T=2 are denoted as $L^{1}_{per}$ and $L^{2}_{per}$. The $\ell_{1}$ loss measures the difference between $\hat{R}$ and $R$ at stage 3 as:
\begin{equation}
\label{equ_2}
L_{\ell_{1}}^{3}=\sum_{m=1}^{H}\sum_{n=1}^{W} \mid \hat{R}(m,n)-R(m,n) \mid,
\end{equation}
where $W$ and $H$ denote the width and height of the image, rspectively.

The total loss is the linear aggregation of the multi-term loss:
\begin{equation}
\label{equ_4}
L_{total}=L^{1}_{per}+L^{2}_{per}+L_{per}^{3}+\lambda L_{\ell_{1}}^{3},
\end{equation}
where $\lambda$ is the weight of $L_{\ell_{1}}^{3}$ to balance different terms.

\subsection{Non-Uniform Illumination Retinal Image Synthesis}
\vspace{0.3cm}
In the real world, it is almost impossible to obtain sufficient paired retinal image with non-uniform illumination and the corresponding clear retinal image. To train the proposed NuI-Go, we approximate the non-uniform illumination degradation model of retinal images by combining the domain knowledge of retinal images with the human-lens attenuation and scattering model. With the above-mentioned retinal image degradation model, we simulate large amounts of degraded retinal images for our model training.

Inspired by the human-lens attenuation and scattering model~\cite{Peli1989}, the retinal image degradation can be reduced to a Retinex model \cite{Land1986} when only the illumination effects are considered:
\begin{equation}
\label{equ_5}
I_{c}(m,n)=R_{c}(m,n)\cdot L(m,n), c\in\{r,g,b\},
\end{equation}
where $I$ represents the observed retinal image; ($m$,$n$) denotes the pixel coordinates; $R$ is the reflectance of the retinal image (\ie, the desired result); and $L$ is the flash illumination of the fundus camera. This simplified model can be described as that the observed retinal image $I$ is decomposed into the product of the reflectance $R$ and the illumination mask $L$. In this paper, we expect to achieve the reflectance of retinal image $R$ from the observed non-uniform illumination retinal image $I$. Here, we assume that each channel of the retinal image has the same illumination mask. According to Eq.~(\ref{equ_5}), given a clear retinal image $R$, a non-uniform illumination mask $L$ is needed to generate the non-uniform illumination retinal image $I$.

Concretely, we first transform the retinal image to the CIE Lab color space and search the well-lit regions in the illumination channel by different thresholds. If the pixel values are larger than the current threshold, they are set to 1. Or these pixels serve as the non-uniform illumination pixels by Gamma correction with random $\gamma$ values ranging from 0.1 to 0.5. The purpose is to preserve the original illumination of the well-lit regions and meanwhile make the low-light regions darker as the non-uniform illumination mask. The potential reason is that some regions in the retinal images are seldom affected by non-uniform or poor illumination, such as optic cup region. In this paper, we use five thresholds from 0.1 to 0.5 with the stride of 0.1. We did not find much improvement when setting more thresholds. After that, a coarse non-uniform illumination mask is generated. Considering the retinal images usually have the smoothed and region-wise non-uniform illumination, we smooth the coarse non-uniform illumination mask by simple 8$\times$ down-sampling and 8$\times$ up-sampling operations. Here, other smooth approaches also can be used. Finally, a retinal image with non-uniform illumination can be synthesized by using Eq. (\ref{equ_5}).

Figure~\ref{fig_2} presents an example of the synthetic image, where the synthetic retinal image has realistic non-uniform illumination, especially preserving the regions which are seldom affected by the non-uniform illumination.

\section{Experiments}
%In this part, we first report the details of the training and implementation of our network. Then, experiments on the synthetic and real-world retinal images and application are conducted, respectively.

\vspace{0.3cm}
\subsection{Training and Implementation Details}
\vspace{0.3cm}
We collect 2,500 well-lit retinal images from the publicly available training set of Kaggle's Diabetic Retinopathy Detection Challenge (KDRDC)\footnote[1]{\url{https://www.kaggle.com/c/diabetic-retinopathy-//detection/}} which provides retinal images taken under a variety of imaging conditions. Following the approach mentioned in Sec. 3.3, we generate 12,500 non-uniform illumination retinal images and resize them to a size of 256$\times$256 due to our limited memory. We randomly split these synthetic images into two parts: 10,000 images for training and the rest as a testing dataset (denoted as \textbf{Test A}).

We implement our network with TensorFlow on a PC with an NVIDIA GTX 1080Ti GPU. During training, a batch size of 8 is applied, and the filter weights are initialized by Gaussian distribution. We use ADAM for network optimization, and fix the learning rate to $1e^{-4}$. We compute perceptual loss at layer relu5\_4 of the VGG-19 network. The weight $\lambda$ in loss function is set to 100. The runtime of the NEDRB is 0.088s for an image with a  size of 256$\times$256, which is fast for practical applications.

\subsection{Evaluation of Retinal Image Non-Uniform Illumination Removal}
\vspace{0.3cm}
In this section, we compare the proposed NuI-Go network with the state-of-the-art models on the synthetic and real retinal images. The synthetic retinal images are denoted as \textbf{Test A}, and the real-world retinal images, denoted as \textbf{Test B}, are obtained from the testing set of KDRDC, including 53,576 images taken under a variety of imaging conditions. The compared methods include recent retinal image enhancement method (SGF \cite{Cheng2018}), traditional image enhancement method (CLAHE \cite{Zuiderveld1994}), and state-of-the-art low-light image enhancement method (LIME \cite{Guo2017}).

\begin{figure*}[!ht]
\centering
\centerline{\includegraphics[width=1\textwidth]{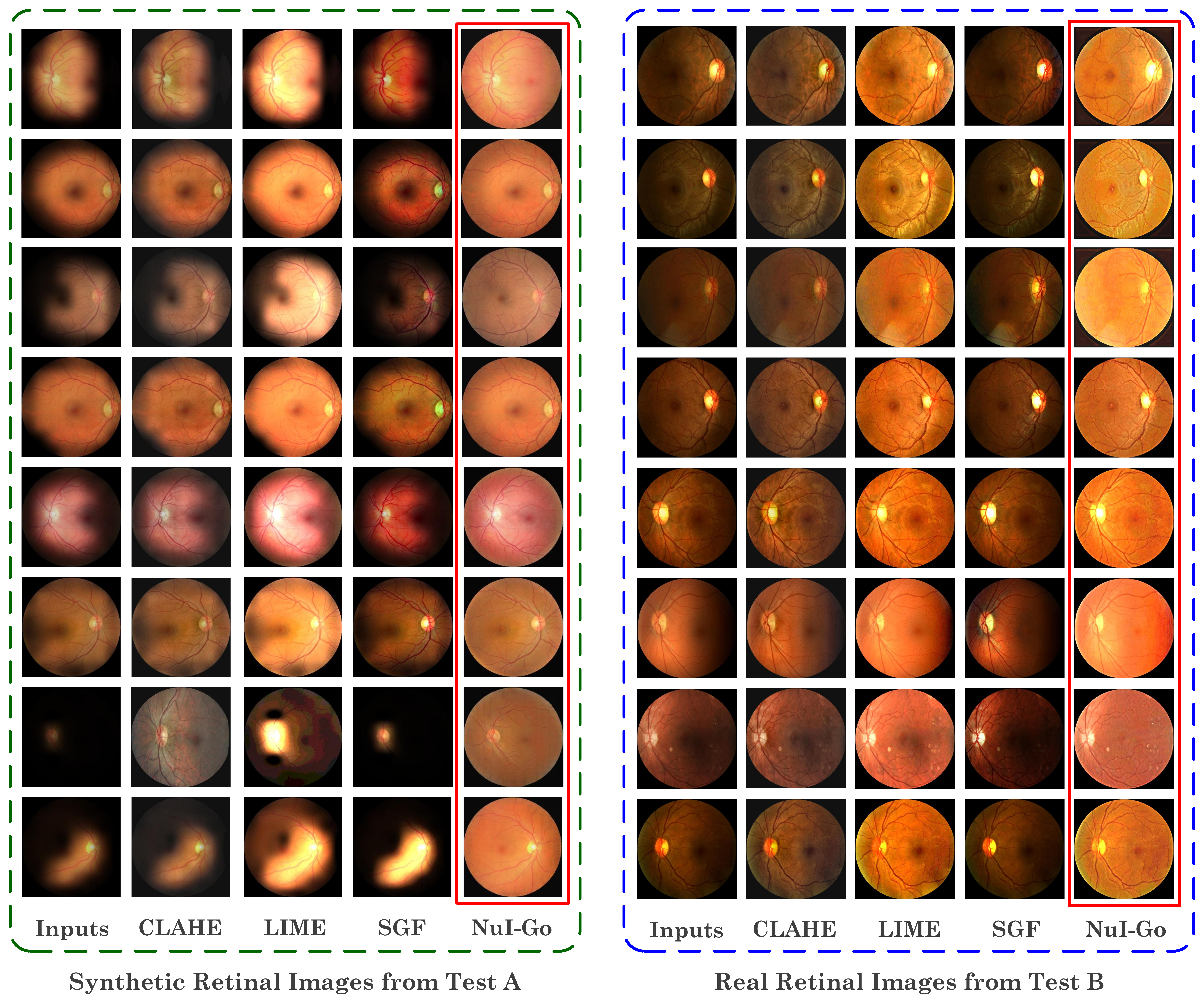}}
\caption{Visual comparisons on synthetic and real retinal images. Red box indicates our results.}
\label{fig_3}
\vspace{0.2cm}
\end{figure*}

The visual comparisons on synthetic and real-world retinal images are presented in Figure~\ref{fig_3}. From it, the methods of CLAHE \cite{Zuiderveld1994} and SGF \cite{Cheng2018} have little effect on the non-uniform illumination regions. LIME \cite{Guo2017} method improves the brightness of retinal images; however, there still remains some poor illumination regions and over-exposured regions on the results due to its limited generalization capability for fundus images. Such results may lead to a clinical misdiagnosis. In contrast, the proposed NuI-Go can effectively correction the non-uniform illumination and preserve the details and natural look of retinal images.  

%\begin{figure*}[!ht]
%\centering
%\centerline{\includegraphics[width=0.6\textwidth]{supply_synthetic.pdf}}
%\caption{Visual comparisons on synthetic retinal images. From left to right are the inputs from \textbf{Test A}, the results of CLAHE \cite{Zuiderveld1994}, LIME \cite{Guo2017}, SGF \cite{Cheng2018}, and our NuI-Go. Red box indicates our results.}
%\label{fig1}
%\end{figure*}
%
%\begin{figure*}[!ht]
%\centering
%\centerline{\includegraphics[width=13cm,height=15cm]{supply_real.pdf}}
%\caption{Visual comparisons on real retinal images. From left to right are the inputs from \textbf{Test B}, the results of CLAHE \cite{Zuiderveld1994}, LIME \cite{Guo2017}, SGF \cite{Cheng2018}, and our NuI-Go. Red box indicates our results.}
%\label{fig2}
%\end{figure*}

We further use the full-reference image quality assessment metrics PSNR and SSIM \cite{Wang2004} for quantitative evaluations on \textbf{Test A}. A higher PSNR value indicates the similarity in terms of pixel-wise values, and a larger SSIM score indicates the result that is closer to the ground truth in terms of structural properties. The average quantitative scores on \textbf{Test A} are reported in Table \ref{tab1}. As presented in Table \ref{tab1}, the proposed NuI-Go achieves the best performance across all metrics. In addition, our method has overwhelming advantages than all the compared methods (CLAHE \cite{Zuiderveld1994}, LIME \cite{Guo2017}, SGF \cite{Cheng2018}) in terms of the average quantitative scores on the synthetic dataset. For example, the average PSNR value of our method is 32.2668 while the second best value (\ie, the result of LIME \cite{Guo2017}) is only 16.3671. For the average SSIM value, it has a similar tendency to that of PSNR. Such a result indicates the effectiveness of the proposed non-uniform illumination removal method.

For \textbf{Test B}, we employ a commonly used non-reference image quality assessment algorithm NIQE \cite{Mittal2013} to assess the perceptual quality. A smaller NIQE score indicates better perceptual quality. Besides, we also conduct a user study to provide realistic feedback about subjective quality. We first randomly select 50 images from \textbf{Test B}, and then invite 10 participants with image enhancement expertise to rank the results of different methods from 1 to 5, where 1 is the worst quality and 5 corresponds to the best quality. This user study is repeated five times. The average values in terms of NIQE \cite{Mittal2013} and user study are shown in Table \ref{tab2}.

\begin{table}
\caption{Quantitative evaluations of PSNR (dB) and SSIM on \textbf{Test A}. For each case, the best result is in bold.}
\centering
\label{tab1}
\setlength{\tabcolsep}{0.6mm}{
\begin{tabular}{l|l|l|l|l|l}
\hline
Metrics &  Inputs & CLAHE \cite{Zuiderveld1994}  & LIME \cite{Guo2017} & SGF \cite{Cheng2018} & NuI-Go\\
\hline
PSNR & 14.6994 & 16.3671 & 14.3072 & 12.2262 &\textbf{32.2668} \\
SSIM & 0.6450  & 0.5065 & 0.6902 &0.5558 &\textbf{0.8281} \\
\hline
\end{tabular}}
\end{table}

\begin{figure*}[!htb]
	\centering
	\includegraphics[width=0.85\textwidth]{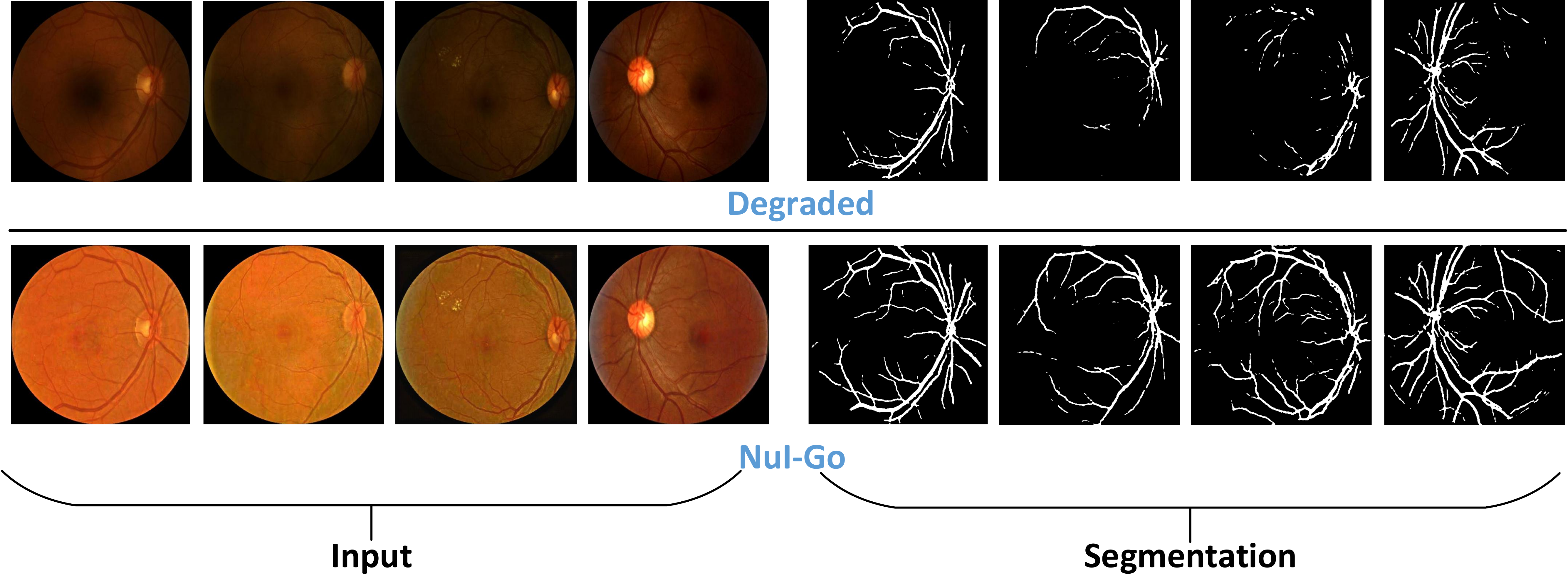}
	\caption{Visual enhancement and retinal vessel segmentation results. \textbf{Left side}: from top to bottom are the degraded retinal images via the proposed non-uniform illumination synthesis method by using the real images sampled from DRIVE dataset \cite{DRIVE} and STARE dataset \cite{STARE1}  and the corresponding enhanced images via the proposed NuI-Go network, respectively. \textbf{Right side}: from top to bottom are the segmentation results of the corresponding degraded images and the segmentation results of the enhanced images by the NuI-Go network, respectively. The segmentation results are produced by the CE-Net \cite{Gu2019}.}
	\label{fig_4}
	\vspace{0.1cm}
\end{figure*}

\begin{table}
\caption{The average non-reference NIQE scores and user study scores on \textbf{Test B}. For each case, the best result is in bold.}
\centering
\label{tab2}
\setlength{\tabcolsep}{0.5mm}{
\begin{tabular}{l|l|l|l|l|l}
\hline
Metrics &  Inputs & CLAHE \cite{Zuiderveld1994}  & LIME \cite{Guo2017} & SGF \cite{Cheng2018} & NuI-Go\\
\hline
NIQE & 7.8087& 6.9539& 7.5331& 8.8697 & \textbf{6.9110}\\
User Study & 2.1452 & 2.5708& 3.2416& 2.4122& \textbf{3.8558} \\
\hline
\end{tabular}}
\end{table}
In Table~\ref{tab2}, our method achieves the best performance in terms of non-reference NIQE \cite{Mittal2013} and subjective user study. Such results demonstrate our method can generalize to real retinal images and effectively improve the visual quality of degraded retinal images. Besides, it also indicates the reasonability of our non-uniform illumination synthesis approach. Though our model is trained on synthetic training data, it is also useful for real retinal images.

In addition, we found our model can be applied to other image data. It is a general model when the corresponding paired training data are available.  However, our network structure is more suitable for retinal image that shares structured texture and similar color. These characteristics of retinal image provide a strong prior for correcting the non-uniform illumination regions by the context.

\begin{table*}
\caption{Quantitative evaluations on DRIVE dataset/STARE dataset in terms of Acc and AUC. For each case, the best result is in bold.}
\centering
\label{tab3}
\begin{tabular}{l|l|l|l|l|l}
\hline
Metrics &  Degraded Inputs & CLAHE \cite{Zuiderveld1994}  & LIME \cite{Guo2017} & SGF \cite{Cheng2018} & NuI-Go\\
\hline
Acc &0.9508/0.9369  & 0.9497/0.9331 & 0.9518/0.9384 &  0.9382/0.9322&\textbf{0.9543}/\textbf{0.9417} \\
AUC &0.9329/0.7976 & 0.9570/0.8509  & 0.9746/0.9076 & 0.9074/0.8014 &\textbf{0.9756}/\textbf{0.9213} \\
\hline
\end{tabular}
\vspace{0.2cm}
\end{table*}

%We apply the proposed NuI-Go as a pre-processing for retinal vessel segmentation.
\subsection{Extended Application of Retinal Vessel Segmentation}  
\vspace{0.3cm}
To evaluate the performance of our method in practical application such as retinal vessel segmentation, we use recent retinal vessel segmentation network CE-Net \cite{Gu2019} which is trained on DRIVE dataset \cite{DRIVE} as a basic retinal vessel segmentation model. We found that the existing vessel segmentation datasets do not contain the non-uniform or poor illumination retinal images. However, the non-uniform illumination retinal images are ubiquitous in the real world. Therefore, we degrade the testing set of DRIVE dataset \cite{DRIVE} and the STARE dataset \cite{STARE1} by the approach proposed by our non-uniform illumination synthesis algorithm. The testing data did not appear in the training set. The first manual annotation in these two datasets is used as the ground truth for performance evaluation. We enhance these testing images by different methods and feed the results to the CE-Net \cite{Gu2019}. At last, we measure the average accuracy (Acc) and the area under receiver operation characteristic curve (AUC) by the formula mentioned in \cite{Gu2019}. First, we present several sample results before and after processed by our method and the corresponding segmentation results in Figure \ref{fig_4}. Then, we report the Acc and AUC scores of different methods in Table~\ref{tab3}.
\begin{figure}[!ht]
	\centering
	\centerline{\includegraphics[width=7.3cm,height=9cm]{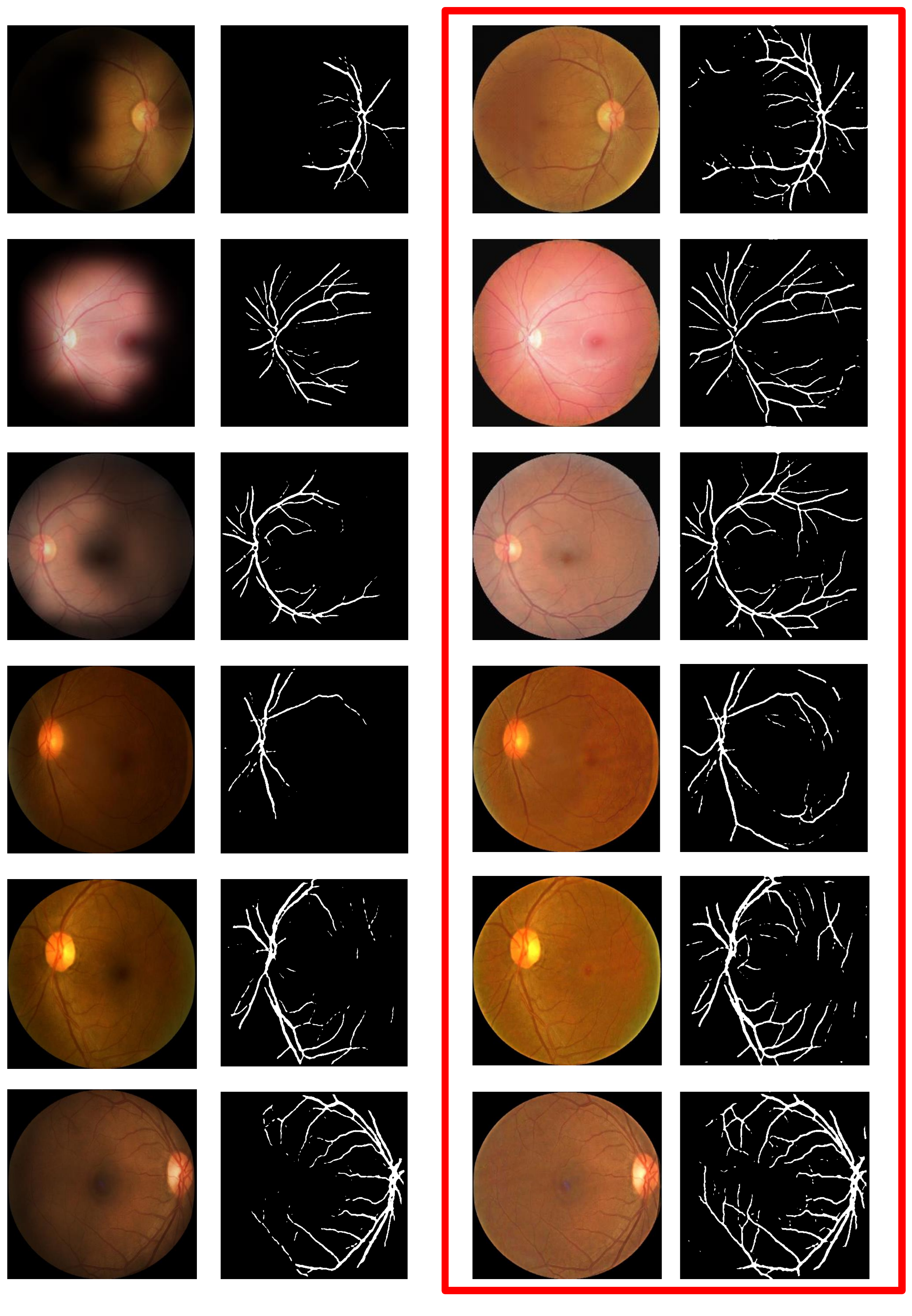}}
	\caption{Visual retinal vessel segmentation results of real retinal images. \textbf{Left side}: the input images from \textbf{Test B} and the corresponding segmentation results. \textbf{Right side}: our enhanced results and the corresponding segmentation results. The segmentation results are produced by the CE-Net \cite{Gu2019}. Red box indicates our results.}
	\label{fig3}
\end{figure}

In Figure~\ref{fig_4}, our method can effectively unveil the vessel under dark regions. As a result, the vessel of retinal images processed by our method can be easily segmented by the CE-Net \cite{Gu2019} which is trained by using well-lit retinal images. Thus, despite the retinal image segmentation algorithms did not consider the non-uniform illumination issue in their designs and training procedures, our method could be used as an effective pre-processing step for further improving their performance.
As presented in Table \ref{tab3}, the results enhanced by the proposed NuI-Go achieve better segmentation performance on the DRIVE \cite{DRIVE} and STARE \cite{STARE1} datasets. Compared with the degraded image inputs, the Acc increases from 0.9508/0.9369 to 0.9543/0.9417, and the AUC increases from 0.9329/0.7976 to 0.9756/0.9213, respectively. This fully illustrates the positive effect of the proposed NuI-Go enhancement method on retinal vessel segmentation task. In addition, it is interesting that some enhancement algorithms decrease the performance of retinal vessel segmentation, such as SGF method. The unstable results are undesired in practical diagnose applications.

To further demonstrate the robustness of our method on real retinal images, we provide several sample segmentation results for real retinal images in Figure~\ref{fig3}. It is obvious that after enhancing by our method, some retinal vessels hidden in dark are recovered and thus can be accurately segmented by the retinal vessel segmentation network CE-Net \cite{Gu2019}. The results in Figure~\ref{fig3} further demonstrate the effectiveness and robustness of our method, which is significant for practical applications.

\vspace{0.2cm}
\section{Conclusion}
\vspace{0.2cm}
In this paper, we propose a deep learning-based retinal image non-uniform illumination removal method. The non-uniform illumination is progressively removed via recursive residual learning while the details and color of retinal images are recovered by the combination of local and non-local features. In addition, to drive the training of the proposed NuI-Go network, we propose a non-uniform illumination degradation model of retinal image by combining the domain knowledge of retinal images with the human-lens attenuation and scattering model. This model can simulate amounts of degraded retinal images with non-uniform illumination and also can be used as a guide for full-reference image quality assessment, which calls for the development of retinal image enhancement. Experiments on synthetic and real-world retinal images show that the proposed method outperforms existing image enhancement methods. Additionally, application experiments suggest that our method can effectively improve retinal vessel segmentation, which is important for practical diagnose application.


\begin{acks}
This work was supported by the National Key Research and Development Program of China (Grant No. 2017YFC0820601), the National Natural Science Foundation of China (Grant No. 61720106004), the Fundamental Research Funds for the Central Universities  (Grant No. 2019RC039),
and the China Postdoctoral Science Foundation (Grant No. 2019M660438).
\end{acks}

%\clearpage

\bibliographystyle{ACM-Reference-Format}
\bibliography{sample-base}

%\appendix

\end{document}